\documentclass[11pt,a4paper, twocolumn]{scrartcl}

\usepackage{casa_conf}	
\usepackage{graphicx}	
\usepackage{blindtext}
\usepackage{hyperref}

\title{DanceGraph: A Complementary Architecture for Synchronous Dancing Online}

\author{David Sinclair\\
        \small \textit{Edinburgh Napier University}\\
        \vspace{-0.2cm}\small D.Sinclair@napier.ac.uk\\
        \and
        Adeyemi Ademola\\
        \small \textit{Edinburgh Napier University}\\
        \vspace{-0.2cm}\small adeyemi.ademola@napier.ac.uk\\
        \vspace{0.1cm}
         \and
        Babis Koniaris\\
        \small \textit{Edinburgh Napier University}\\
        \vspace{-0.2cm}\small B.Koniaris@napier.ac.uk\\
        \and
        Kenny Mitchell\\
        \small \textit{Roblox and Edinburgh Napier University}\\
        \vspace{-0.2cm}\small k.mitchell2@napier.ac.uk\\}

\begin{document}

\maketitle

\begin{abstract}
DanceGraph is an architecture for synchronized online dancing overcoming the latency of networked body pose sharing. We break down this challenge by developing a real-time bandwidth-efficient architecture to minimize lag and reduce the timeframe of required motion prediction for synchronization with the music's rhythm. In addition, we show an interactive method for the parameterized stylization of dance motions for rhythmic dance using online dance correctives.
\end{abstract}
\linebreak
\linebreak

\keywords{Dance, Latency, Motion Capture, Animation, Network Architecture}


\section{Introduction}

DanceGraph addresses the challenge of achieving synchronized online dancing against the latency inherent in the transmission of network signals among remote dance partners. We use a divide-and-conquer approach of firstly developing an efficient architecture for the lowest possible lag of interactive immersive experiences \cite{lag1993mackenzie}. Secondly, our system designed for lag reduction, virtuously contributes to lower durations of required motion prediction necessary for assured synchronization according to the rhythm of the dance's music. Therefore, we present two primary networked dance latency counter-measures,
\begin{itemize}
    \item Our engine-agnostic DanceGraph architecture that achieves reduced end-to-end latency by providing \emph{as-direct-as-possible} paths from sensors to display
    \item Real-time dance correctives applied as rhythmic motion predictions to present multiple body motions synchronized with local music for each online dance partner's experience
\end{itemize}

We offer a high-performance technique for reducing the bandwidth required for real-time animation in a lossless manner, utilizing data-driven bounds specifically tailored for use in synchronous online dancing. An additional outcome of our online dance correctives is a parameterized stylizing of dance motions for rhythmic dance. Finally, we assess implementation alternatives for networked avatar frameworks including high-performance network transport layers in our open-source DanceGraph solution.

\section{Related Work}

Numerous research studies have delved into the realm of human motion synthesis \cite{butepage_deep_2017,yan_2019_ICCV, habibie_recurrent_2017}, with a focus on understanding the relationship between different modalities, such as dance and music, for improved sophistication of motion synthesis. These works are related to our methods in the sense that motion synthesis can be developed for the purpose of motion prediction in reducing latency in online dancing. Initially, in motion synthesis, statistical models have been prevalent, and the motion segments were synthesized by selecting from a database that matched the features of the music segment, such as rhythm, structure, and intensity \cite{shiratori_dancing--music_2006, fan_example-based_2012}.

More recently, deep learning approaches include, ChoreoMaster \cite{chen_choreomaster_2021} which employs an embedding module to capture the music-dance connection, and San et al's DeepDance \cite{sun_deepdance_2021} is a cross-modal association system that correlates dance motion with music. Lee et al. \cite{zhuang_music2dance_2020} present a decomposition-to-composition framework that utilizes MM-GAN for music-based dance unit organization. Yang et al \cite{YangEtAl} utilize normalizing flow techniques to learn the probability distribution of dance motions, which are conditioned on input music beats and artist-generated keyframes, so as to interpolate plausible dance motions between the keyframes. Aristidou et al \cite{Aristidou} breaks dance motions into intermediate-sized components, called motifs, with temporal length of roughly two beats; these are synthesized and combined with features derived from the musical spectrum and concatenated into a higher-level dance choreography. Pan et al \cite{pan2022diverse} uses a complex set of encoders and a dataset made of dance motions to allow fine-grained control of generated motion sequences. In DanceFormer \cite{li_symbiotic_2021}, a kinematics-enhanced transformer-guided networks were used for motion curve regression while in DanceNet \cite{zhuang_music2dance_2020}, a musical context-aware encoder fuses music and motion features. Recently, cross-modal transformers such as, Transflower \cite{valle-perez_transflower_2021}, have been successfully applied to model the distribution between music and motion. 

Wang et al. \cite{wang2022} introduce a novel task of Music-driven Group Dance Synthesis, where the goal is to synthesize group dance with style collaboration. To address this challenging task, the work constructs a rich-annotated 3D Multi-Dancer Choreography dataset (MDC) and proposes a framework called GroupDancer, consisting of three stages: Dancer Collaboration, Motion Choreography, and Motion Transition. To make the model trainable and able to synthesize group dance with style collaboration, the authors propose mixed training and selective updating strategies. The proposed GroupDancer model is evaluated on the MDC dataset and shown to synthesize satisfactory group dance synthesis results with style collaboration. The goal of music-conditioned dance synthesis is to generate dance motion sequences that are associated with a given musical context. Here, they also apply music-conditioned dance synthesis, but instead for the purpose of latency correction of synchronous online dance experiences.

Recent excellent work in motion and shape capture from sparse markers (MoSh) \cite{Loper:SIGASIA:2014} synthesizes convincing full-body soft tissue reproductions of body motions from sparse pose inputs. With Loper et al's lifelike non-rigid human body poses, we apply our novel synchronous dance correctives of sparse body poses in real-time for realistic avatar visualization.

\begin{figure*}[h]
\centering
\includegraphics[width=1\textwidth]{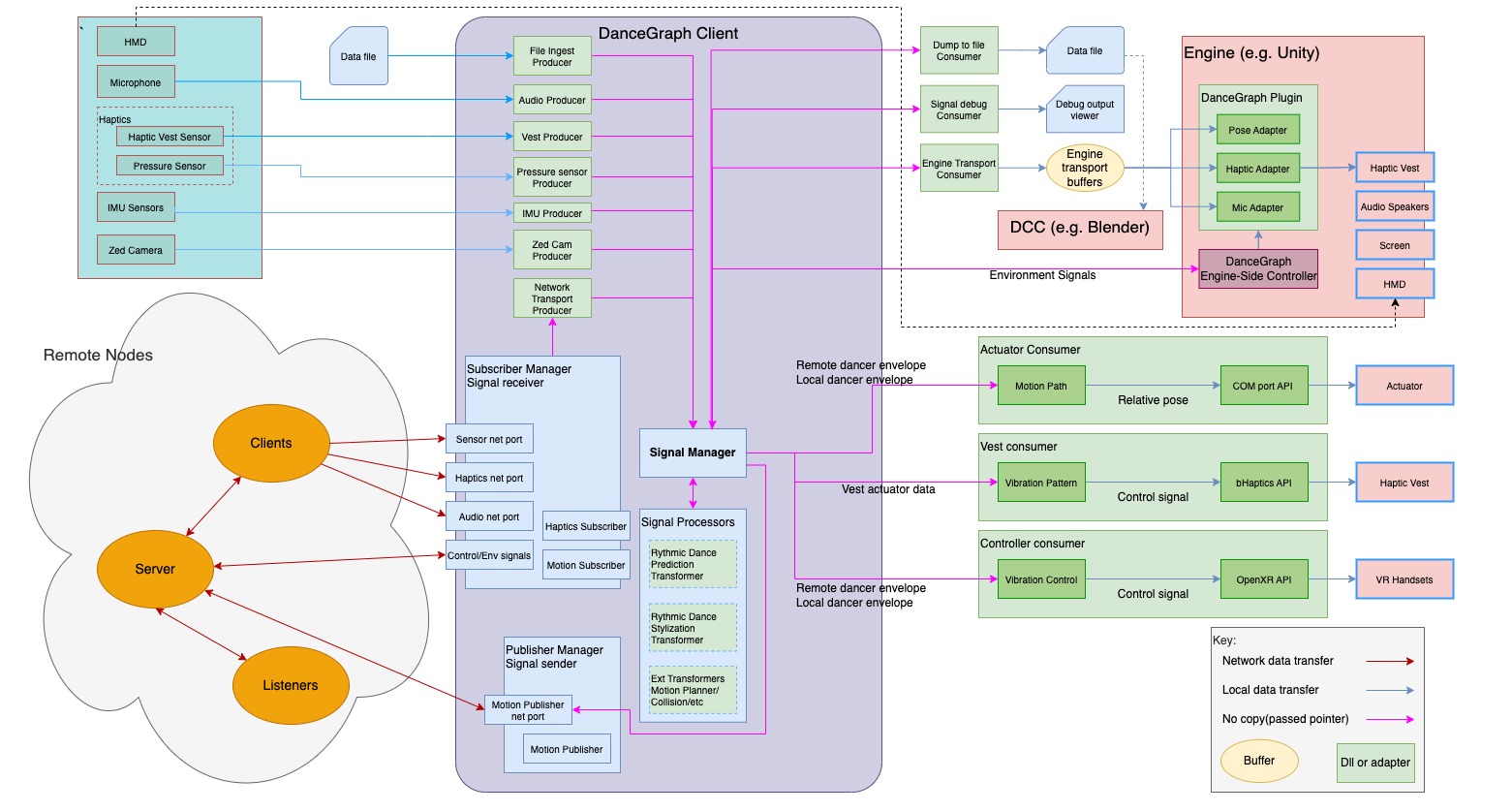}
\caption{A DanceGraph client receives sensor inputs locally from that guest's devices along with remote guests' signals (via the Network Transport Producer) and applies (Signal Manager) composed outputs \emph{as directly as possible} to attached displays and actuators. That is, clients intercept hardware sensor signals before they arrive at media applications circumventing the latency of buffered engine updates.}
\label{fig:dancegraph}
\end{figure*}

\section{\emph{As-Direct-as-Possible} Networked Dance Architecture}

\emph{DanceGraph} is a low-latency networked architecture for online dancing and avatar-driven social interactions, with a driving principle of \emph{as-direct-as-possible} digital paths between input sensors and output displays and actuators. The architecture comprises flexible modular specifications of event-driven processing \cite{michelson2006event} with input sensor and output actuator producers and consumers processing live streams of multi-modal data, selectively published and subscribed to \cite{gagliardi1995} according to the access of each guest's available devices and compute power. 

Integration with virtual worlds is realized through existing game engines, all connected and communicating in a client-server framework. However, each output device can be driven directly from DanceGraph, e.g., for fast response of haptic actuators, or from within connected adapter components of the attached game engine. These elements are built around fast real-time rendering display output with tuned integration of virtual reality head-mounted display (HMD) motion-adapted rendering.

And so, the key enabler of lower latency in the architectural design of DanceGraph is this intercept of signals before they arrive in the client software update loop. Signals from local sensor hardware are sent to the client application at the same time as they are sent to the network so that client software-related sources of latency are avoided. (Figure \ref{fig:dancegraph}). DanceGraph is open source\footnote{\url{https://github.com/CarouselDancing/DanceGraph}} and extensible with custom proprietary components or further open source services providing access to new sensors and devices with adapters for Unity and launcher application for independent DanceGraph servers.

\subsection{Decoupled and Engine-Agnostic}

Online interactive media applications such as networked dancing experiences typically use \emph{off-the-shelf} game engines such as Unity or Unreal Engine to reduce the implementation effort, using considerably-optimized components for rendering, physics, AI, and more. Sensory hardware vendors frequently supply libraries that can connect with such game engines so that developers can write logic based on sensor data retrieved or applied during the engine's frame update step.

A significant, but preventable, source of latency is polling the data \emph{and} performing networking logic at the rate of the often broadly-purposed game engine's update step. 
Ensuring synchronization between a local signal and a remote avatar representation involves multiple steps, including polling the local signal, sending it to the networking subsystem, transmitting it over the network, and updating the remote client application. However, these steps add to the overall latency, which can be further compounded by the structure of the game engine and the buffering of data. Therefore, DanceGraph is intentionally engine agnostic and decoupled by means of \emph{adapters} connecting the chosen client engine, which may be implemented as native plugins, inter-process communication (IPC), or message passing and streaming. 

The signal, in order to be synchronized with the remote avatar representation, needs to be polled in the local application, then sent to the networking subsystem, which will then send the data to the remote client (most likely at a future update step due to buffering or execution order), which will be transmitted over the network to the remote client application, which will read that data during its own update step and inform the client avatar at that or a future update step. A number of update steps are executed due to the game engine structure, and each step adds to the total latency. Despite recent mechanisms such as \emph{low latency frame syncing} \cite{ue427docs2021} game engines can still exhibit as much as 120ms of lag over and above network lag.

\subsection{Producer-Consumers and Adaptors}

The DanceGraph architecture is implemented primarily in C++ to enable a portable and low-latency, high-performance approach to communications between sensors, applications, and other clients across the network.

Network signals are passed into DanceGraph through `producer' modules (typically these are hardware drivers or network listeners), and are distributed via the Signal Manager onto 'consumer' modules (typical uses for these are to pass signals onward to the network, or to pass them to a game engine or DCC for viewing, or saving data to a file). Consumers are preferably written to be agnostic to the type of signal data being consumed. A producer’s signal may be routed to multiple consumers. In the current implementation, real-time motion capture is performed by the sensor hardware and SDK, with the use of a ZED\footnote{https://www.stereolabs.com/zed-2/} camera, and the Dancegraph producer module merely performs the task of passing the tracking data to the client. Producers can be written either to extract tracking information from other suitably-functional cameras, or alternatively, to calculate the tracked motion themselves from raw camera data. The actual routing of the data is performed by the signal manager, a relatively simple portion of the architecture, which keeps track of which producer signals are linked to which consumers directly by means of pointers to in-memory data structures, allocated by the manager itself, in addition to optionally associating metadata as necessary.

In order to view sensor data incoming from consumers, the architecture provides for a native language plugin, e.g. a C++ plugin for the Unity game engine, which receives data in a shared memory interprocess communication or other data transfer mechanism. These signal Adapter modules, such as the Pose Adaptor, embedded within the game engine, convert or marshal the received signal data into data structures suitable for rendering using the game engine's typical scripting mechanisms.

\subsection{Bandwidth Reduction}
In the context of real-time online synchronous dance little time is afforded to compression of motion signals, therefore in this work, we focus on low-cost quantization of quaternion pose joint orientations. Each joint of our sparse human avatar rig was analyzed independently across a range of dance motions to efficiently define per joint and quaternion component's numerical ranges, in order to determine compression with minimal loss of data. Furthermore, because of the redundancy in the quaternion representation, one of the quaternion components could be discarded and reconstructed from the other three; for this particular use-case, the `w' component of an (x, y, z, w) quaternion was almost always the largest, and therefore the best candidate to be dropped, negating the need to use two bits per quaternion to indicate the dropped component.

\begin{figure*}
\centering
\includegraphics[width=.75\textwidth]{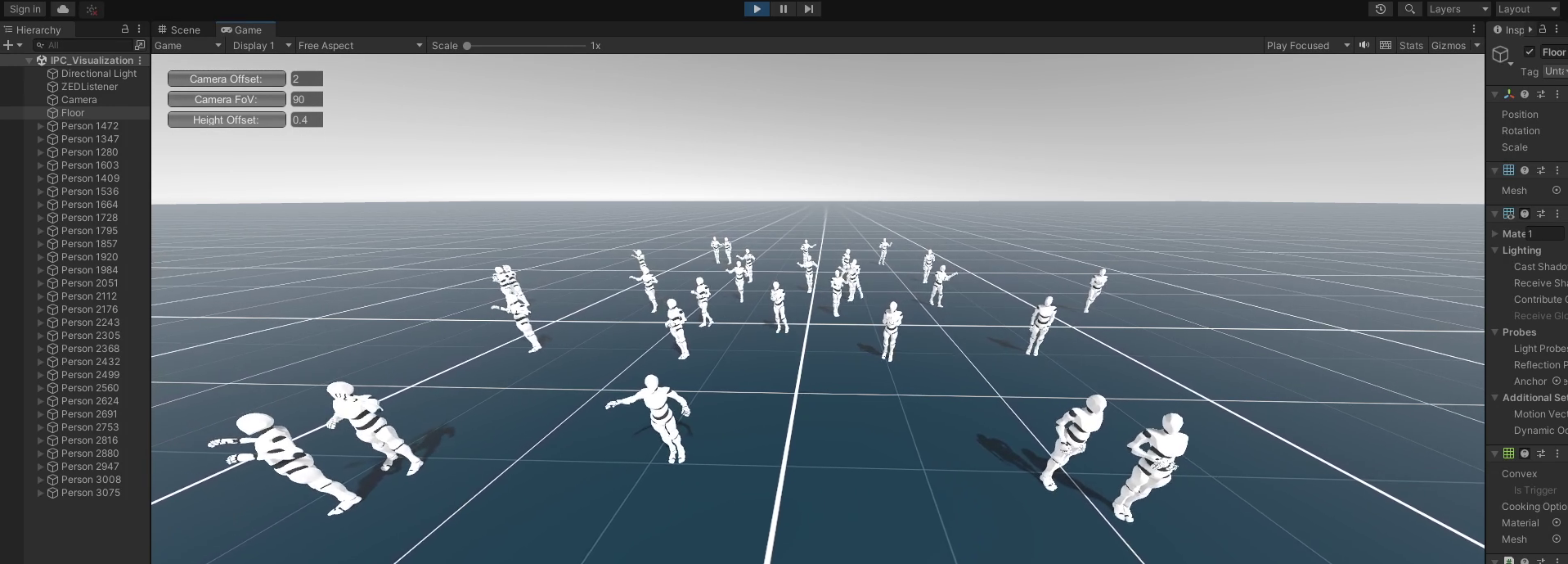}
\caption{Thirty connected simulated dancers, generated using prerecorded motion signals transmitted among DanceGraph clients, the signals being simulated are indistinguishable from those generated in real-time by live ZED camera captured dancers.}
\label{fig:dancers30}
\end{figure*}

This lossless approach reduced overall motion data signal transmissions to two-thirds of the original network data stream size. A perceptually unnoticeable oriented lossy compression scheme would be expected to yield further bandwidth reductions. However, this was already sufficient to achieve 30 simultaneous online dance motion signals with our Unity adapter implementation of DanceGraph (see figure \ref{fig:dancers30}).

\subsection{Measured Latency Reductions}

We demonstrate latency savings when compared with similar networked applications. For the first two of these comparisons, we compared latency with a regular dance networking implementation, using the ZED SDK for Unity, and the Mirror\footnote{https://mirror-networking.gitbook.io/docs/trivia/a-history-of-mirror} networking system with the \emph{SyncVars} method of synchronizing each dancers' pose state.

\subsubsection{Bypass of Engine Networking}
By intercepting the signals from the hardware and passing them directly upstream to the network, we gain the most significant latency savings. Here, packets are queued to the network immediately upon being processed by the ZED native C++ SDK, rather than via ZED's Unity C\# plugin interface.

In our tests, the regular Unity method was able to acquire the packet in the engine's C\# scripting language 79.3ms after the ZED camera image of the footage was retrieved from the device into RAM. In Unity then, it was queued upstream, using the \emph{SyncVars} method provided by the Mirror networking library. By contrast, DanceGraph is able to queue the tracking data to the network within 16.5ms on average from the baseline footage. This results in an average saving of 62.8ms, short-cutting both the Unity update loop and game engine networking infrastructure. We note the bulk of the remaining time is spent solving the tracking pose from the camera image within the native ZED SDK and a further source of latency optimization opportunity.

\subsubsection{Ablation Measure of Bypassing Engine Update Delays}
As well as offering significant latency savings by circumventing regular game engine networking infrastructure, our \emph{direct as possible} offers large savings when measured isolated from networking and viewing only locally generated signals. Here, the signal is sent to the game engine on the local machine for viewing in the engine.

Again, the regular Unity method takes about 79.3ms until the signal is available to the engine's scripting facilities. With our method, the signal reached Unity 24.8ms after the camera image was retrieved into RAM, on average, therefore our DanceGraph architecture gives a latency saving of 54.5ms independently of networking and only subject to engine update loop architecture delays.

\subsubsection{Optimized Network Buffering}
One further saving is in the native DanceGraph's bespoke network buffering infrastructure in comparison to that of more flexible implementations. The Unity Transport Package implementation, as implemented in their 'Boss Room' demo, gives a round trip networking time of over 16ms.

Here, the DanceGraph system is able to route the network infrastructure from the client to the server to another client in just under 10ms, on average, saving 3.01ms in network buffering. This saving is in addition to that described in the above section on engine networking bypass.

\begin{figure}
    \centering
    \includegraphics[width=0.45\columnwidth]{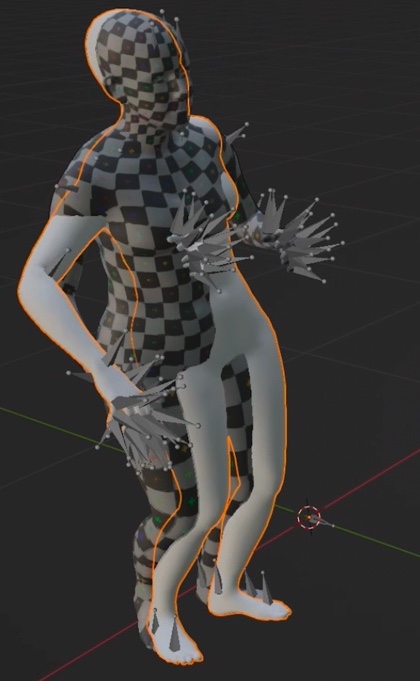}
    \includegraphics[width=0.45\columnwidth]{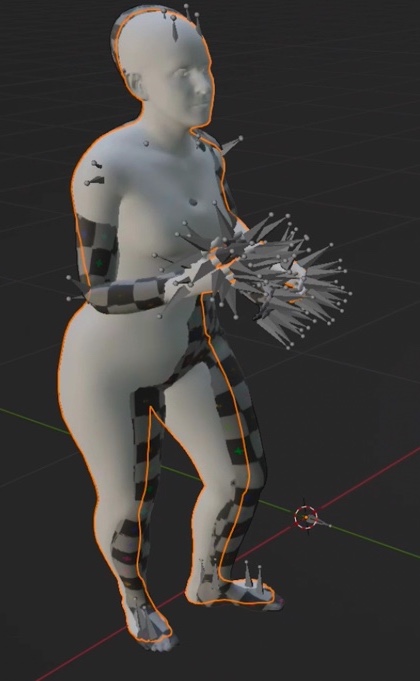}
    \caption{Feature-based dance stylization and motion amplification in Blender at 30fps. The checked model is the original dancing avatar, and the grey model is our amplified stylized avatar motion showing wider hip sway motions.}
    \label{fig:DanceAmplify}
\end{figure}

\section{Rhythmic Motion Prediction for Synchronous Online Dance}
Whilst the above steps can be used to significantly reduce the latency due to client software, the physical limitation of time a digital signal takes to transmit from one remote location to another in addition to the regular latency of external networking infrastructure is inevitable. To counter this hard constraint, we employ a synthetic motion correction in the \emph{Rythmic Dance Prediction Transformer} (Figure \ref{fig:dancegraph}) to alter the incoming pose sequences according to the rhythmic motion features of the dance's local music time context so that the remote dancers appear locally to have synchronous 'in-time' motions with the musical beat.

For example, in the \emph{50020-salsa-1-stageii} salsa dance sample from the MoSh dataset\cite{Loper:SIGASIA:2014} (also part of AMASS\cite{AMASS:ICCV:2019}) we identify the dominant periods (or periodogram) of the time-series of joint motions of the dance, e.g. the extremities of hip-sway motions, and the remap these detected rhythmic motion features in time according to the beat of the musical sequence. 

Since we drive motion temporal corrections on the sparse avatar rig skeletal motion into the MoSh\cite{Loper:SIGASIA:2014} IK framework, resulting motions are physically plausible accordingly. Further, deep learning advances for sparse rig-to-mesh motion models can be applied similarly.

As a natural extension of our dance motion feature detection, we can simply amplify the periodic motions to yield exaggerated and emphasized dance motions of avatars (see figure \ref{fig:DanceAmplify}) with controlled parameterization of body motion zones such as hips, shoulders, hands, etc. 

\section{Conclusion}

We have proposed a novel architecture that tackles the challenge of achieving synchronized online dancing between remote dance partners. Our approach breaks down the problem by designing a bandwidth-efficient architecture that directly minimizes lag and reduces the duration of motion prediction needed for synchronization with the music's rhythm. We provide a survey of motion synthesis works to set the context for our online dance synchronization with our real-time prediction method.

Additionally, dance rhythmic synchronization for networked dancers with prediction provides robustness to long-duration hitches in lag between partners, occasional but common connection lapses, and affords a new framework for live dance styling. We introduce such a method for the parameterized stylization of dance motions that leverages our synchronous online dance correctives. Our findings pave the way for more effective solutions in achieving robust online dancing bringing geographically dispersed dance partners closer together than before.
 
\section*{Acknowledgements}
This work is supported by funding EU’s Horizon 2020 research and innovation programme under grant agreement No. 101017779.


\bibliographystyle{plain}
\bibliography{quaternionsAndMotionPredictions}

\begin{thebibliography}{10}

\bibitem{Aristidou}
Andreas Aristidou, Anastasios Yiannakides, Kfir Aberman, Daniel Cohen-Or, Ariel
  Shamir, and Yiorgos Chrysanthou.
\newblock Rhythm is a dancer: Music-driven motion synthesis with global
  structure, 11 2021.

\bibitem{butepage_deep_2017}
Judith Butepage, Michael~J. Black, Danica Kragic, and Hedvig Kjellstrom.
\newblock Deep {Representation} {Learning} for {Human} {Motion} {Prediction}
  and {Classification}.
\newblock pages 6158--6166, 2017.

\bibitem{chen_choreomaster_2021}
Kang Chen, Zhipeng Tan, Jin Lei, Song-Hai Zhang, Yuan-Chen Guo, Weidong Zhang,
  and Shi-Min Hu.
\newblock {ChoreoMaster}: choreography-oriented music-driven dance synthesis.
\newblock {\em ACM Transactions on Graphics}, 40(4):145:1--145:13, July 2021.

\bibitem{fan_example-based_2012}
Rukun Fan, Songhua Xu, and Weidong Geng.
\newblock Example-{Based} {Automatic} {Music}-{Driven} {Conventional} {Dance}
  {Motion} {Synthesis}.
\newblock {\em IEEE Transactions on Visualization and Computer Graphics},
  18(3):501--515, March 2012.
\newblock Conference Name: IEEE Transactions on Visualization and Computer
  Graphics.

\bibitem{gagliardi1995}
M.~Gagliardi, R.~Rajkumar, and L.~Sha.
\newblock The real-time publisher/subscriber inter-process communication model
  for distributed real-time systems: design and implementation.
\newblock In {\em 2013 IEEE 19th Real-Time and Embedded Technology and
  Applications Symposium (RTAS)}, page~66, Los Alamitos, CA, USA, may 1995.
  IEEE Computer Society.

\bibitem{ue427docs2021}
Epic Games.
\newblock {\em UE4.27 - Unreal Engine 4 Documentation - Low Latency Frame
  Syncing}.
\newblock Epic Games.

\bibitem{habibie_recurrent_2017}
Ikhsanul Habibie, Daniel Holden, Jonathan Schwarz, Joe Yearsley, and Taku
  Komura.
\newblock {\em A {Recurrent} {Variational} {Autoencoder} for {Human} {Motion}
  {Synthesis}}.
\newblock January 2017.

\bibitem{li_symbiotic_2021}
Maosen Li, Siheng Chen, Xu~Chen, Ya~Zhang, Yanfeng Wang, and Qi~Tian.
\newblock Symbiotic graph neural networks for 3d skeleton-based human action
  recognition and motion prediction.
\newblock {\em IEEE Transactions on Pattern Analysis and Machine Intelligence},
  44(6):3316--3333, 2021.
\newblock Publisher: IEEE.

\bibitem{Loper:SIGASIA:2014}
Matthew~M. Loper, Naureen Mahmood, and Michael~J. Black.
\newblock {MoSh}: Motion and shape capture from sparse markers.
\newblock {\em ACM Transactions on Graphics, (Proc. SIGGRAPH Asia)},
  33(6):220:1--220:13, November 2014.

\bibitem{lag1993mackenzie}
I.~Scott MacKenzie and Colin Ware.
\newblock Lag as a determinant of human performance in interactive systems.
\newblock In {\em Proceedings of the INTERACT '93 and CHI '93 Conference on
  Human Factors in Computing Systems}, CHI '93, page 488–493, New York, NY,
  USA, 1993. Association for Computing Machinery.

\bibitem{AMASS:ICCV:2019}
Naureen Mahmood, Nima Ghorbani, Nikolaus~F. Troje, Gerard Pons-Moll, and
  Michael~J. Black.
\newblock {AMASS}: Archive of motion capture as surface shapes.
\newblock In {\em International Conference on Computer Vision}, pages
  5442--5451, October 2019.

\bibitem{michelson2006event}
Brenda~M Michelson.
\newblock Event-driven architecture overview.
\newblock {\em Patricia Seybold Group}, 2(12):10--1571, 2006.

\bibitem{pan2022diverse}
Junjun Pan, Siyuan Wang, Junxuan Bai, and Ju~Dai.
\newblock Diverse dance synthesis via keyframes with transformer controllers,
  2022.

\bibitem{shiratori_dancing--music_2006}
Takaaki Shiratori, Atsushi Nakazawa, and Katsushi Ikeuchi.
\newblock Dancing-to-{Music} {Character} {Animation}.
\newblock {\em Computer Graphics Forum}, 25(3):449--458, 2006.
\newblock \_eprint:
  https://onlinelibrary.wiley.com/doi/pdf/10.1111/j.1467-8659.2006.00964.x.

\bibitem{sun_deepdance_2021}
Guofei Sun, Yongkang Wong, Zhiyong Cheng, Mohan~S. Kankanhalli, Weidong Geng,
  and Xiangdong Li.
\newblock {DeepDance}: {Music}-to-{Dance} {Motion} {Choreography} {With}
  {Adversarial} {Learning}.
\newblock {\em IEEE Transactions on Multimedia}, 23:497--509, 2021.
\newblock Conference Name: IEEE Transactions on Multimedia.

\bibitem{valle-perez_transflower_2021}
Guillermo Valle-Pérez, Gustav~Eje Henter, Jonas Beskow, André Holzapfel,
  Pierre-Yves Oudeyer, and Simon Alexanderson.
\newblock Transflower: probabilistic autoregressive dance generation with
  multimodal attention.
\newblock {\em ACM Transactions on Graphics}, 40(6):1--14, December 2021.
\newblock arXiv:2106.13871 [cs, eess].

\bibitem{wang2022}
Zixuan Wang, Jia Jia, Haozhe Wu, Junliang Xing, Jinghe Cai, Fanbo Meng, Guowen
  Chen, and Yanfeng Wang.
\newblock Groupdancer: Music to multi-people dance synthesis with style
  collaboration.
\newblock In {\em Proceedings of the 30th ACM International Conference on
  Multimedia}, MM '22, page 1138–1146, New York, NY, USA, 2022. Association
  for Computing Machinery.

\bibitem{yan_2019_ICCV}
Sijie Yan, Zhizhong Li, Yuanjun Xiong, Huahan Yan, and Dahua Lin.
\newblock Convolutional sequence generation for skeleton-based action
  synthesis.
\newblock In {\em Proceedings of the IEEE/CVF International Conference on
  Computer Vision (ICCV)}, October 2019.

\bibitem{YangEtAl}
Zhipeng Yang, Yu-Hui Wen, Shu-Yu Chen, Xiao Liu, Yuan Gao, Yong-Jin Liu, Lin
  Gao, and Hongbo Fu.
\newblock Keyframe control of music-driven 3d dance generation.
\newblock {\em IEEE Transactions on Visualization and Computer Graphics}, pages
  1--12, 2023.

\bibitem{zhuang_music2dance_2020}
Wenlin Zhuang, Congyi Wang, Siyu Xia, Jinxiang Chai, and Yangang Wang.
\newblock {Music2Dance}: {DanceNet} for {Music}-driven {Dance} {Generation},
  March 2020.
\newblock arXiv:2002.03761 [cs, eess].

\end{thebibliography}

\end{document}